\documentclass[referee]{mn2e}
\usepackage{graphics}
\title{On the formation and abundance of CO in envelopes of AGB stars}

\author[R. Papoular]{R. Papoular$^{1}$\thanks{E-mail:
papoular@wanadoo.fr}\\
$^{1}$Service d'Astrophysique and Service de Chimie Moleculaire,\\
CEA Saclay, 91191 Gif-s-Yvette, France}
\begin{document}

\date{Accepted . Received ; in original form }

\pagerange{\pageref{firstpage}--\pageref{lastpage}} \pubyear{2002}
   \maketitle
\label{firstpage}
\begin{abstract}
It is generally considered, as a rule of thumb, that carbon monoxide forms very early in envelopes of AGB stars, and that it consumes most of the carbon, or most of the oxygen, depending on whether  the photosphere is oxygen-rich or carbon-rich, respectively. \rm This work focuses on the latter case , with the purpose of quantifying the remaining fraction of gaseous carbon which is then available for forming carbonaceous grains. Since AGB stars are (probably) the main providers of cosmic carbon grains, this residual fraction is essential in establishing the validity of current grain models. Here, we use a kinetic treatment to follow the chemical evolution of circumstellar shells towards steady state. It is shown that the residual fraction depends essentially on the atomic ratio of pristine gaseous carbon and oxygen, and on the cross-section for CH formation by collision of C and H atoms. It lies between 55 and 144 C atoms per $10^{6}$ H atoms, depending on the values adopted for unknown reaction rates and cosmic C abundance. This is much larger than predicted by the rule of thumb recalled above.

The present results depend strongly on the rate of the reaction C+H$\rightarrow$CH: far from thermodynamic equilibrium (which is the case here), CO cannot be formed if this rate is as low as generally assumed. We have, therefore, estimated this rate by chemically modelling the reaction and found it indeed much higher, and high enough to yield CO abundances compatible with observations. An accurate experimental rate determination is highly desirable.

  \end{abstract}

\section{Introduction}
The spectral analysis of the CS (CircumStellar) envelopes of AGB stars has been an active field of observational astronomy for decades (see, for instance, Bujarrabal et al. \cite{buj}, Woods et al. \cite{woo}). As a result, the list of detected atoms, ions, radicals and molecules is continuously enriched and their characterization improves in extent and quality as instruments and telescopes are constantly upgraded (see van Dishoek \cite{vdi} for references).

One of the most abundant molecules in both O-rich and C-rich AGBs is carbon monoxide, CO, so much so that the dominant isotopic variety, $^{12}$CO, is optically thick in its characteristic infrared bands, precluding direct determination of its column density towards the star. One way of circumventing this difficulty is to measure instead the density of the rarer isotope, $^{13}$CO, from which that of $^{12}$CO is deduced, assuming their ratio is somewhere between 1/50 and 1/100, because of (undetermined) fractionation processes which locally alter the relative cosmic abundance of 1/89 (see Burgh at al. \cite{bur}, Liszt \cite{lis}).

In spite of these achievements and because of observational uncertainties (such as the CO isotopic ratio) and the relative scarcity of stars amenable to in-depth analysis, it is still difficult to draw a complete and accurate observational picture of the variations of CO abundance with the type of AGB star. 

For want of a better guess, one may assume an arbitrary but reasonable relative abundance, such as $[CO]/[H_{2}]=10^{-3}$ (e.g. Woods \cite{woo})). Another simple rule of thumb has often been used: all the oxygen (respectively carbon) expelled by C-rich (respectively O-rich) stars is locked into gaseous CO. This rule is strongly suggested by the high reciprocal chemical affinities of the two elements, and the strength of the CO bond.

Precisely because of the latter, the carbon and oxygen atoms locked in CO are irreversibly lost for carbonacious and silicate grains which ultimately condense in the envelope. If strictly applied, the rule above implies that only stars with C/O$>$1 can contribute to carbon grains. Now, for many carbon grain models, the C budget is already quite tight, so the available C input deserves careful scrutiny. In particular, the carrier of the extinction bump at 2175 $\AA{\ }$ (see, for instance, Bless and Savage \cite{ble}) requires large quantities of pure carbon (see Snow and Witt \cite{sno} for a discussion). Since CS shells, together with supernovae and novae, where grain condensation probably occurs in similar ways, are the main, if not sole, genitors of grains, accurate and  generally applicable  procedures are required to estimate the amount of C available for grain condensation.

 On the theoretical side, a parallel effort was made to model the chemistry which takes place in these envelopes, using long chains of known processes and reactions. Assuming local thermodynamical equilibrium (LTE) has been a powerful approach for estimating the relative abundances of the various species of interest as a function of temperature, pressure and distance from the central star (see Tsuji \cite{tsu}, Salpeter \cite{sal}, Duley and Williams \cite{dul}). However, for LTE to be applicable, steady states at high densities and temperatures are required, so that detailed balance may be invoked, which is not the case in CS. Besides, it is not easy to include solid grains in the LTE scheme.

 Another approach consists in selecting a number of molecular species and a number of known reactions between these species, and expressing the rate of change of each species as a function of the instantaneous densities of all species and the rates of their reactions. In general, a steady state is quickly established, which depends on local temperature and initial densities. The complexity of the problem is illustrated by the work of Prasad and Huntress \cite{pra}, a milestone in the study of the chemistry of intersellar molecular clouds. While the physical conditions in clouds are not comparable with those of CS envelopes, that work is an inspiring example for our present purposes. Cherchneff \cite{che06} recently used the kinetic approach to study the dependance of molecular populations in the inner parts of CS of AGB stars, as a function of the C/O ratio in the photosphere. The present research is also based on chemical kinetics rather than LTE, but the goal is much less ambitious, being limited to estimating the fraction of carbon available for grain formation.

\rm We consider carbon grains and gaseous CO to be competitors for the available C atoms from the photosphere. A small number of chemical reactions between the most abundant gaseous chemical species are retained to define the two corresponding routes for pristine C atoms (Sec. 2). They translate into a set of differential equations which can be solved numerically for given relative abundances of C and O, to yield the relative number of C atoms that end up in grains and CO respectively (Sec. 3). The results are arranged as a function of the ratio C/O, displaying continuous trends from O-rich to C-rich stars. The fraction of carbon available for grains is thus predicted to be overall much larger than present estimates (Sec. 4).

The discussion of these results (Sec. 5) singles out the reactions which dominate the production of CO or carbonaceous grains, and should therefore be more fully documented by laboratory measurements. Finally, we discuss how the average ratio of carbon atoms available for grains to total H atoms can be predicted based on the present computations.

\section{The chemical reactions}
Although its bonding energy is particularly strong (250 kcal/mol or about 11 eV), CO is not known to form directly by simple association of neutral C and O atoms. This is analogous to the case of H$_{2}$: because of the similarity of the two interacting atoms, conservation of energy and momentum requires the presence of a third body (grains or high pressure gas). This is not available in the environments considered here. The emission of a molecular resonance photon of 11 eV could help, but no such resonance is tabulated (see Radzig and Smirnov \cite{rad}). We therefore look for indirect paths, involving successive binary reactions.

The number of possible reactions in the CS gas is huge (cf Prasad and Huntress \cite{pra}, UMIST Data Base \cite{let}), even for the small number of atomic species that are involved initially: H, C and O. For present purposes, it is necessary and sufficient to select a few of them that could be considered as essential or sufficiently representative of the complete process. We now set out to do this.

The gas densities and temperatures to be considered are characteristic of inner regions of AGB envelopes illuminated by photospheres at 4000 K or less, and shielded from the ionizing UV photons of the IS medium. Neutral radicals and molecules will therefore be dominant over ions, as suggested by thermodynamic equilibrium models (Tsuji \cite{tsu}, Duley and Williams \cite{dul}). Of course, cosmic rays pervade CS envelopes, but their rate of ionization of matter (e.g. $10^{-17}$ s$^{-1}$ per H atom) is negligible, considering the rates of chemical reactions to be retained below
 (in the order of $10^{-4}$ s$^{-1}$).

Of the most abundant elements, helium will be disregarded because it is chemically inactive, as will be nitrogen because it does not enter in grains in quantities large enough to compete with the processes considered here.

The initial reservoir of the atomic species at the start of calculations will therefore consist of H, C and O. Their absolute densities only affect the time constants of the reactions, not their outcome, which is determined by their relative densities. The H density will be set at $10^{10}$ cm$^{-3}$ for all computed cases, which corresponds to a mass loss rate of about $3\,\,10^{-5}$ M$_{\odot}$/y, for a wind velocity of 10 km.s$^{-1}$, at a radial distance of $10^{14}$ cm from the center of the star. Given the very low relative abundance of the other elements, this H density is not notably altered by the chemical evolution of the gas, and may therefore be assumed to remain constant, for the sake of simplicity.

 In selecting possible reactants, preference will be given to those carrying smaller number of atoms of higher cosmic abundance. O$_{2}$ is excluded on the grounds that, in this case, the balance between formation and destruction processes is highly unfavourable to its presence in significant quantities.

Kinetic and LTE calculations as well as observations show that the abundance of gaseous molecules with more than 3 or 4 atoms is much lower than that of CO (see Duley and Williams \cite {dul}). Besides, the products of chemical reactions between carbonaceous reactants with more than  3 atoms are either light carbon-poor molecules (mainly CO), or heavier hydrocarbon molecules. This is because C can form rings or long chains, while O cannot. We therefore limit our chemical scheme to reactants with less than 4 atoms each, and assume that products with 2 or more C atoms will further react to end up mainly in carbon grains. This assumption is borne out by our calculations below. For instance, in Fig. 2, the only carbon-bearing molecules that survive to any significant extent are CO and  C$_{2}$H$_{2}$. Reaction between O and C$_{2}$H$_{2}$ may still yield CO, but its rate is negligible (Rate06, No 321). Other pathways from C$_{2}$H$_{2}$ have been considered by Cherchneff et al. (\cite {che92}; e.g. to C$_{4}$H$_{4}$ through the vinylidene radical C$_{2}$HH), however, they found that, ultimately, C$_{2}$H$_{2}$ remained the most abundant molecule from which carbon grains are grown.

\rm Reactions with activation energy higher than 2500K are excluded, as are those with a rate prefactor a$\leq 10^{-11}$ cm$^{3}$s$^{-1}$. In further limiting the number of reactions to be retained among the selected species, we require that one at least of the products also belongs to this selection. Finally, trimolecular reactions are obviously excluded at the relevant gas densities.

These (arbitrary but argumented) conditions limit the number of selected species to 11: H, H$_{2}$, C, C$_{2}$, O, OH, CO, CH, CH$_{2}$, C$_{2}$H and C$_{2}$H$_{2}$. The reactions involved are listed in Table 1 with their particulars as given in the UMIST Data Base, Rate06 \cite{wood}. Interestingly, no way was found to form CO or carbon grains without CH as an intermediary, and the only route to CH, according to the cited data bases, appears to be radiative association of C and H. But, even with prior ionization by cosmic rays, the rate given in Rate06 is extremely low ($10^{-17}$ cm$^{3}$s$^{-1}$). True enough, the rates of reactions 11 (H+C$_{2}\to$CH+C) and 47 (H$_{2}$+C$\to$CH+H) are quite high, but they were excluded because of excessive activation energy (30450 and 11700 K, respectively). We were therefore led to use chemical modelling in order to study the reaction C+H$\to$CH (details in the Appendix). It was found that it has no activation energy, and the cross-section is of order $30\,\AA{\ }^{2}$ in the temperature range of interest. The same procedure was applied to the association of CH with CH, although this is not as crucial as the former reaction. 

\begin{table*}
\caption[]{Reaction scheme; k(cm$^{3}$s$^{-1}$)=$10^{-10}\,\,$a(T(K)/300)$^{b}$exp(-c/T(K)); for k1, k2, see Appendix.}
\begin{flushleft}
\begin{tabular}{llllll}
\hline
Rate06 No & Reactants & Products & a & b & c (K) \\
\hline
1 & H+CH & C+H2 & 1.3 & 0 & 80 \\
\hline
 2 & H+CH2 & CH+H2 & 0.664& 0 & 0 \\ 
\hline
 48 & H2+CH & CH2+H & 5.46 & 0 & 1943 \\
\hline
 63 & C+CH & C2+H & 0.659 & 0 & 0 \\
\hline
 65 & C+CH2 & C2H+H & 1 & 0 & 0 \\ 
\hline
 73 & C+OH & CO+H & 1 & 0 & 0 \\
\hline
 74 & C+C2H & C3+H & 1 & 0 & 0 \\
\hline
 138 & CH+O & OH+C & 0.254 & 0 & 2381 \\
\hline
140 & CH+O & CO+H & 1.02 & 0 & 914 \\
\hline
237 & CH2+CH2 & C2H2+2H & 1.8 & 0 &400 \\
\hline
241 & CH2+O & CO+2H & 1.33 & 0 & 0 \\
\hline
242 & CH2+O & CO+H2 & 0.8 & 0 & 0 \\
\hline
316 & O+CO2 & CO+C & 6 & 0 & 0 \\
\hline
317 & O+C2H & CO+CH & 0.17 & 0 & 0 \\
\hline
k1 & H+C & CH & 1 & 0 & 0 \\
\hline
k2 & CH+CH & C2H2 & 1 & 0 & 0 \\
\hline
 
\end{tabular}
\end{flushleft}
\end{table*}

\section{The evolutionary equations}
The calculations will show that the transient between initial and steady states is so short that it takes place entirely in a very small radial span. This spatial evolution can therefore be overlooked. Let $n_{H}(0)$, $n_{C}(0)$ and $n_{O}(0)$ be the initial densities, and $\dot{n}$ the time derivative of $n$. Based on the essential constancy of the gas temperature and the small dependance  of the rate constants on temperature, we shall replace each rate constant, k, by its prefactor written as \emph {a} (cm$^{3}$s$^{-1}$), followed by its ranking number in Rate06.

The following equations were solved as difference equations, with a time step dt=$10^{-1}$s.

\begin{eqnarray*}
& &\dot n_{H2}=-a_{48}n_{H2}n_{CH}+a_{1}n_{H}n_{CH}+a_{2}n_{H}n_{CH2}+a_{242}n_{O}n_{CH2};\\
& &\dot n_{C}=-a_{63}n_{C}n_{CH}-a_{73}n_{C}n_{OH}-k_{1}\,n_{C}n_{H}\\
& &-a_{65}n_{C}n_{CH2}+a_{1}n_{H}n_{CH};\\
& &\dot n_{O}=-a_{140}n_{O}n_{CH}-a_{138}n_{O}n_{CH}-a_{317}n_{O}n_{C2H}-a_{316}n_{O}n_{C2}\\
& &-a_{241}n_{O}n_{CH2}-a_{242}n_{O}n_{CH2};\\
& &\dot n_{CO}=a_{73}n_{C}n_{OH}+a_{140}n_{CH}n_{O}+a_{241}n_{CH2}n_{O}+a_{242}n_{CH2}n_{O}\\
& &+a_{316}n_{O}n_{C2};\\
& &\dot n_{CH}=-a_{1}n_{H}n_{CH}-a_{63}n_{C}n_{CH}-a_{138}n_{CH}n_{O}-a_{140}n_{CH}n_{O}\\
& &-2k_{2}n_{CH}n_{CH}-a_{48}n_{H2}n_{CH} +a_{2}n_{H}n_{CH2} +a_{317}n_{C2H}n_{O}+k_{1}n_{C}n_{H};\\
& &\dot n_{OH}=-a_{73}n_{OH}n_{C}+a_{138}n_{CH}n_{O};\\
& &\dot n_{C2}=-a_{316}n_{C2}n_{O}+a_{63}n_{C}n_{CH};\\
& &\dot n_{CH2}=-a_{2}n_{CH2}n_{H}-a_{65}n_{C}n_{CH2}-2a_{237}n_{CH2}n_{CH2}-a_{241}n_{O}n_{CH2}\\
& &a_{242}n_{O}n_{CH2}+a_{48}n_{H2}n_{CH};\\
& &\dot n_{C2H}=-a_{317}n_{C2H}n_{O}-a_{74}n_{C}n_{C2H}+a_{65}n_{C}n_{CH2};\\
& &\dot n_{C2H2}=a_{237}n_{CH2}n_{CH2}+k_{2}n_{CH}n_{CH};\\
& &n_{H}=n_{H}(0)-n_{CH}-n_{C2H}-2n_{CH2}-2n_{C2H2};\\
& &n_{C}=n_{C}(0)-n_{CH}-2n_{C2H}-n_{CH2}-2n_{C2H2};\\
& &n_{O}=n_{O}(0)-n_{CO}-n_{OH};\\
& &n_{Cgr}=2(n_{C2}+n_{C2H}+n_{C2H2}).\\
\end{eqnarray*}

Note that, in order to ensure conservation of atoms, current values of first generation atoms,  $n_{C}$ and $n_{O}$, are not computed by incrementation like the reaction products. The spatial density, $n_{Cgr}$, of C atoms trapped in carbonaceous grains is $\emph {defined}$ as a sum over C-enriched, later generation, species. The equation for $n_{H}$ is only meant to check that it does not vary significantly.

\section{The results}
Several runs were made to cover a significant set of parameters. As an illustration, Fig. 1 shows the time evolution of the densities of H, H$_{2}$, C and O for initial values $n_{C}(0)=n_{O}(0)=3\,\,10^{6}$ and $n_{H}(0)=10^{10}$ cm$^{-3}$. Steady state is essentially reached in about one day, justifying the neglect of the space variable. Molecular hydrogen also forms quickly but does not exceed $\sim 0.1\,n_{H}$.

\begin{figure}
\resizebox{\hsize}{!}{\includegraphics{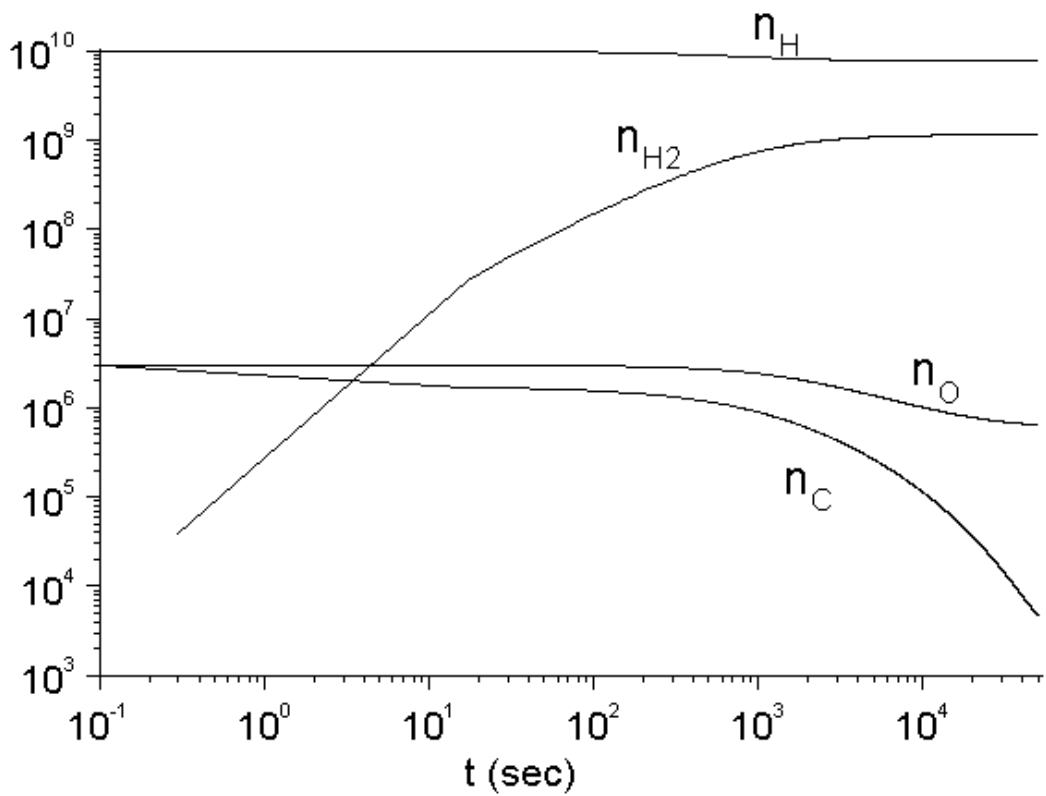}}
\caption[]{A typical example of computational results. The parameters are the initial values of the spatial densities of first generation species (cm$^{-3}$): $n_{H}=10^{10}, n_{C}=n_{O}=3\,\,10^{6}, n_{H2}=0$. Although initially absent, molecular hydrogen evolves quickly, then levels off at a small density relative to H. Carbon is totally consumed, contrary to oxygen. Steady state is essentially reached in a day's time. The hydrogen density decreases only slightly, due to the formation of hydrogenated species.}
\end{figure}

Figures 2 and 3 show that carbon is entirely consumed into CO and grains while oxygen ends up partly in OH and mainly in CO. Figure 2 illustrates the detailed budget of carbon: C$_{2}$ and the primary hydrocarbons, CH, CH$_{2}$, C$_{2}$H quickly decay into C$_{2}$H$_{2}$, which, in this truncated scheme of reactions, is a final product, as is CO. Now, C$_{2}$H$_{2}$ is known to be the seed of carbonaceous grains (see Frenklach and Fiegelson \cite{fre}). This justifies the truncation of the reaction chain for our present purposes.

\begin{figure}
\resizebox{\hsize}{!}{\includegraphics{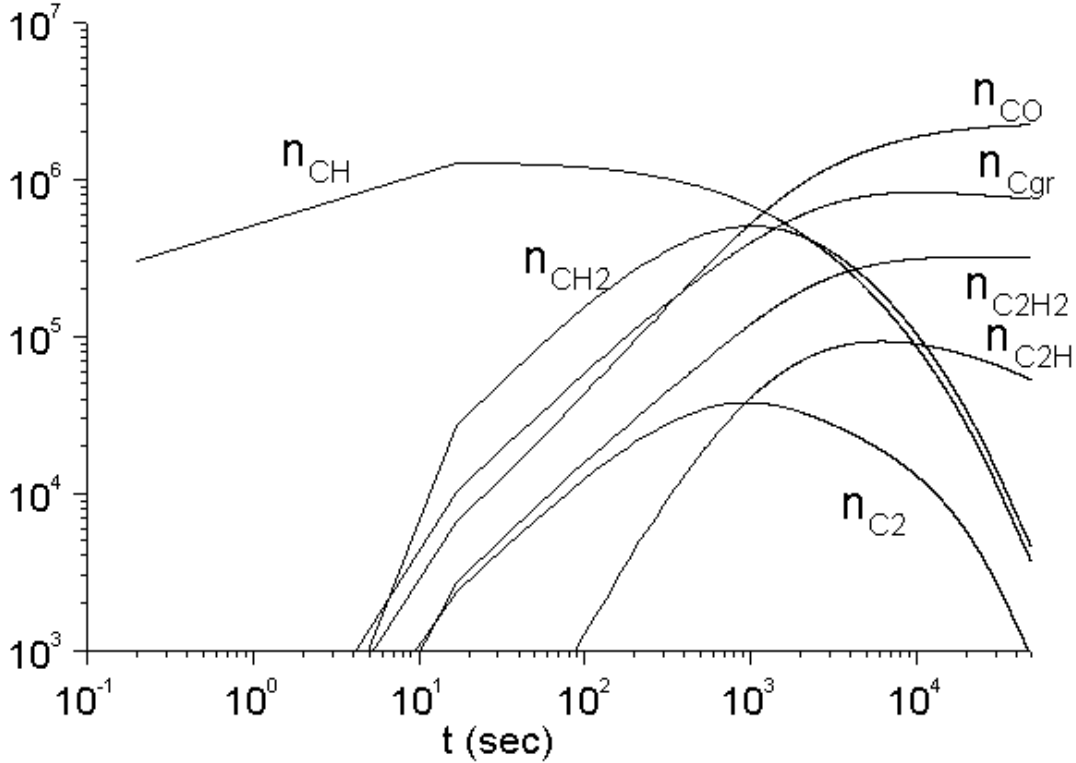}}
\caption[]{Same initial densities as in Fig. 1. Evolution of C-rich species. They all decay to zero, except acetylene. CH is the indispensable starter and mediator of the evolution.}
\end{figure}

\begin{figure}
\resizebox{\hsize}{!}{\includegraphics{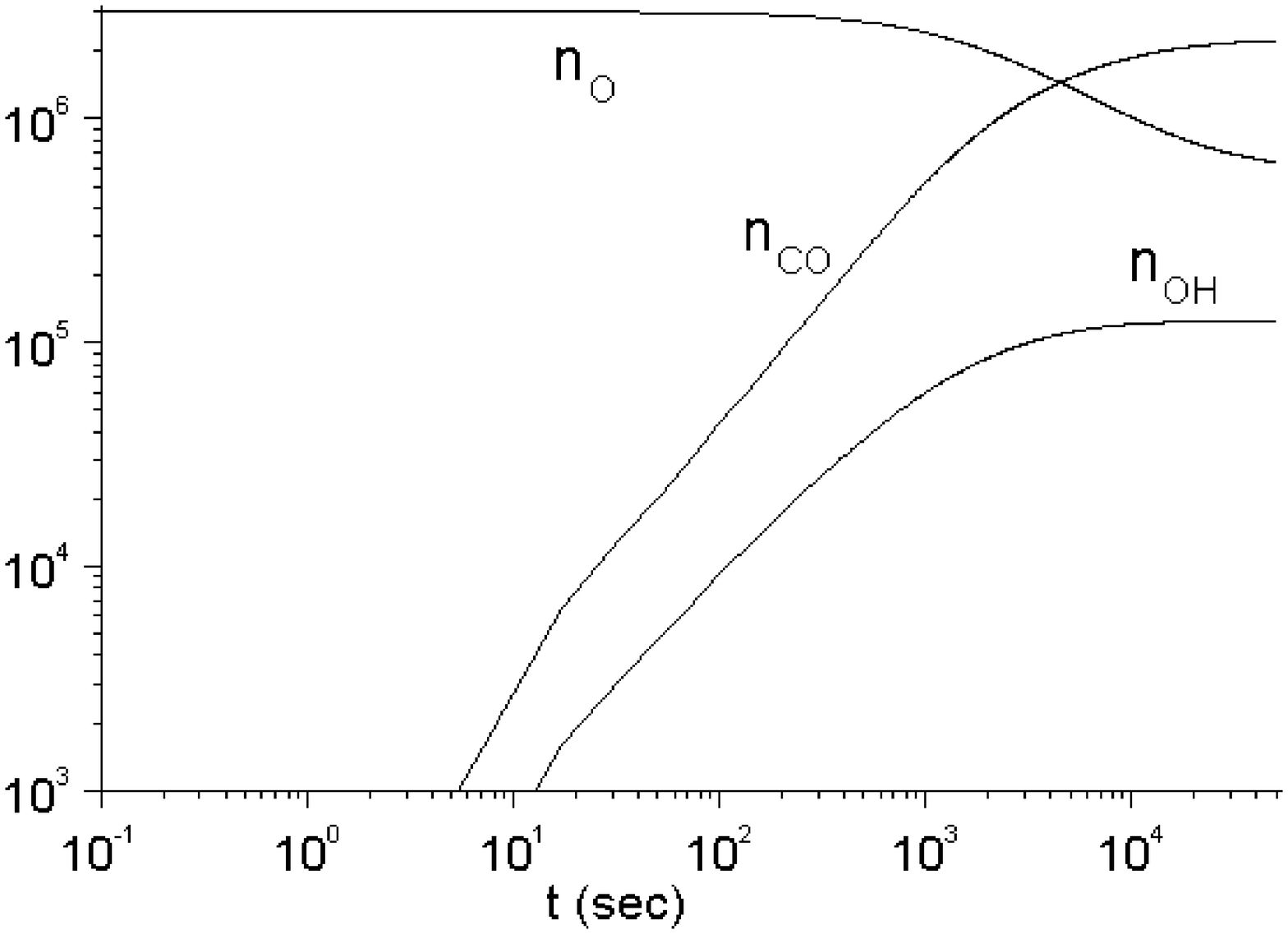}}
\caption[]{Same initial densities as in Fig. 1. Evolution of O-rich species. Oxygen is essentially consumed into CO, and partly OH.}
\end{figure}

Several runs with different initial densities of mother atoms showed that asymptotic values of products, expressed as fractions of $n_{C}(0)$, vary notably with the ratio $n_{C}(0)/n_{O}(0)$, but are quite insensitive to the absolute values of these two densities. The asymptotic values of interest, $n_{CO}(f)/n_{C}(0)$ and $n_{Cgr}(f)/n_{C}(0)$, are therefore plotted in Fig. 4 and 5 as a function of $n_{C}(0)/n_{O}(0)$,  for a few cases spanning the relevant range (filled squares; straight lines are drawn between squares to help the eye). 

\begin{figure}
\resizebox{\hsize}{!}{\includegraphics{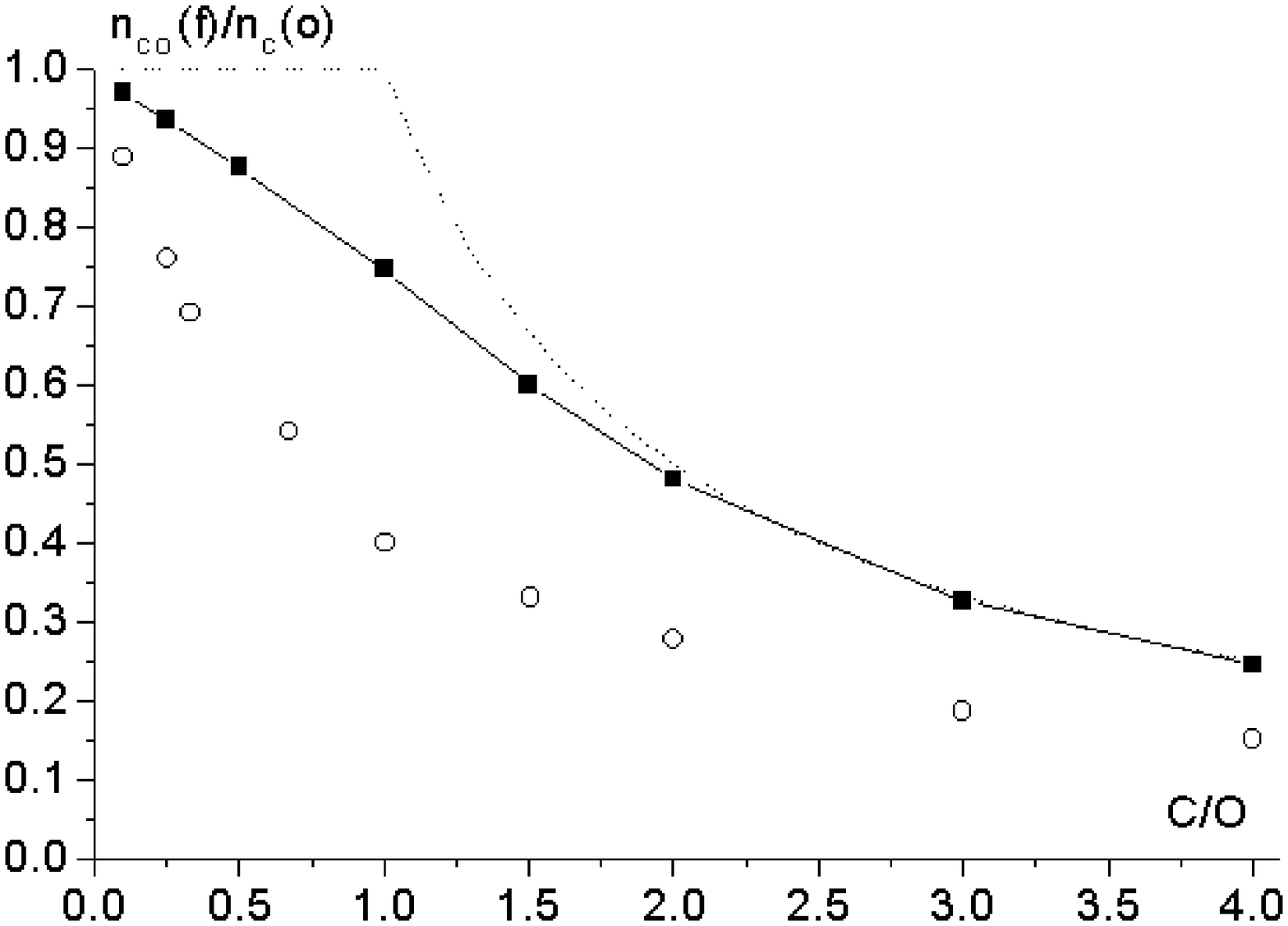}}
\caption[]{The fraction of initial carbon that is consumed in CO, as a function of the initial ratio of C to O (C/O=$n_{C}(0)/n_{O}(0)$; f designates the end of the calculation run). Filled squares: solutions of the set of equations in Sec. 3 (straight lines were drawn between computed points as an aid to the eye). Dashed line: illustration of the usual rule of thumb (see text): it runs through the computed points in the limits of very small and very large C/O, but the deviations are considerable in the intermediate range. Open circles: solutions of the reduced set of reactions  and equations (Sec. 5) ; inspite of the drastic simplification, the general trend is similar to that of the squares, with differences smaller than a factor 2.}
\end{figure}

\begin{figure}
\resizebox{\hsize}{!}{\includegraphics{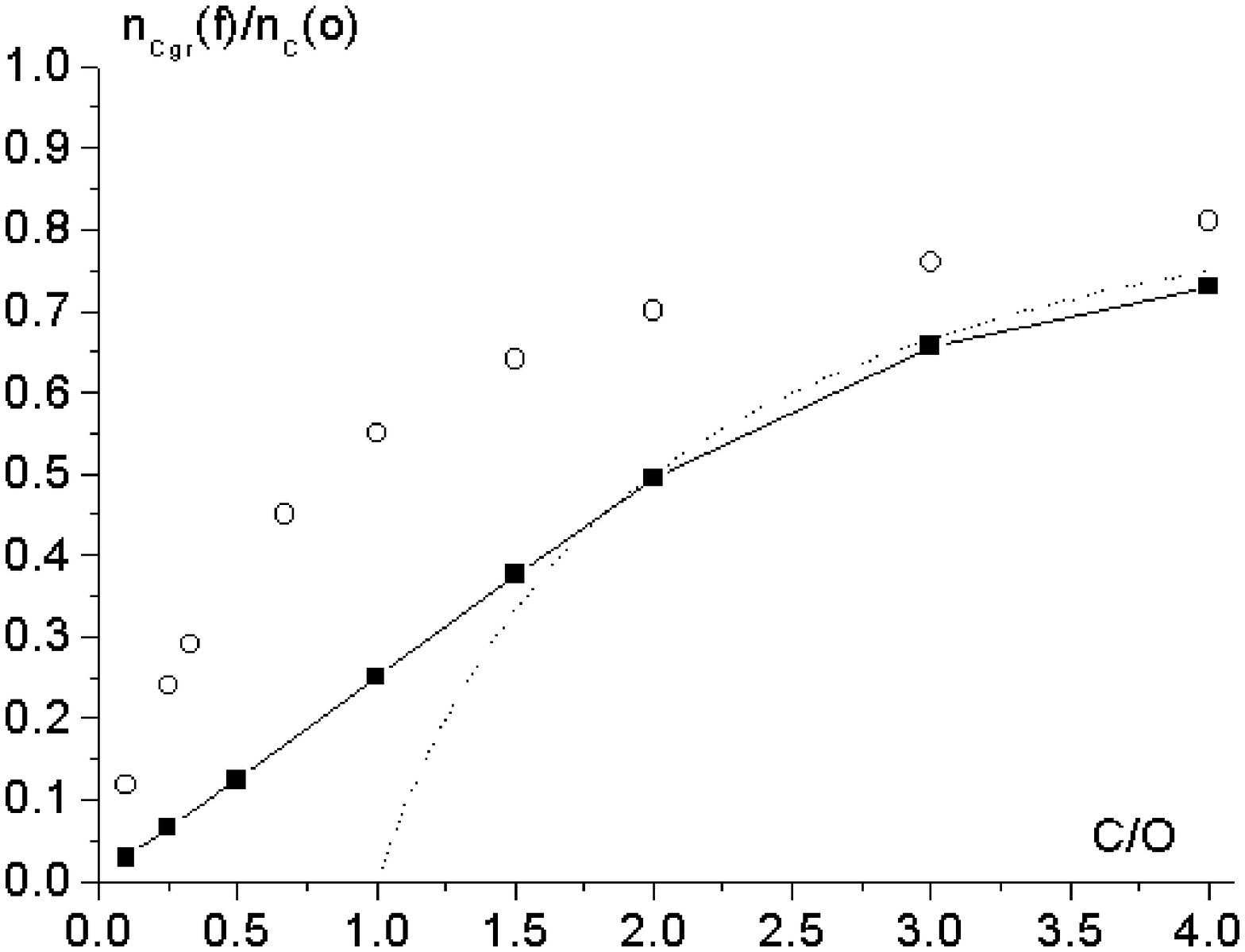}}
\caption[]{The fraction of initial carbon that is ultimately consumed into carbon grains (represented here by C$_{2}$H$_{2}$ essentially). Same symbol conventions as in Fig. 4.}
\end{figure}

It may help to know that a square polynomial fit to the points in Fig. 4 gives, to a very good approximation,

\begin{equation}
\frac {n_{CO}(f)}{n_{C}(0)}=1.018-0.328\,\,\frac {n_{C}(0)}{n_{O}(0)}-0.033\,\,\frac  {n_{C}(0)}{n_{O}(0)}^{2} . 
\end{equation}

In Fig. 5, the filled square with abscissa 4 is lower than the corresponding thumb-rule value. This is abnormal and due to insufficient time length of the run: the curve $n_{Cgr}$=f(t) did not reach its asymptote in this case.

Finally, if one is interested in the fraction of oxygen available for silicate grains, an upper limit for it is $n_{O}(f)/n_{O}(0)$,  which is plotted in Fig. 6.

\section{Discussion}

\begin{figure}
\resizebox{\hsize}{!}{\includegraphics{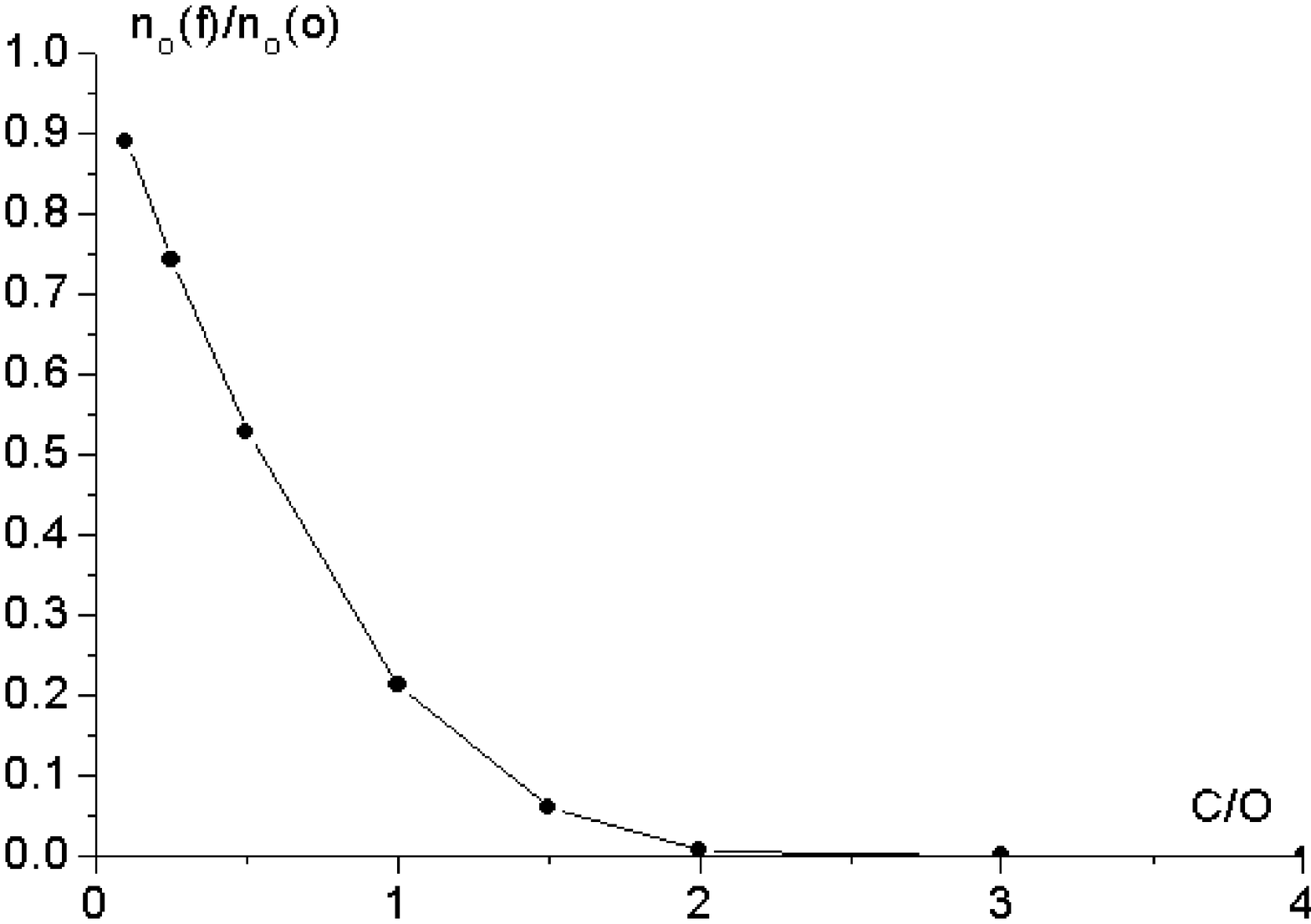}}
\caption[]{An upper limit to the fraction of oxygen available for silicate grains, $n_{O}(f)/n_{O}(0)$.}
\end{figure}

\begin{figure}
\resizebox{\hsize}{!}{\includegraphics{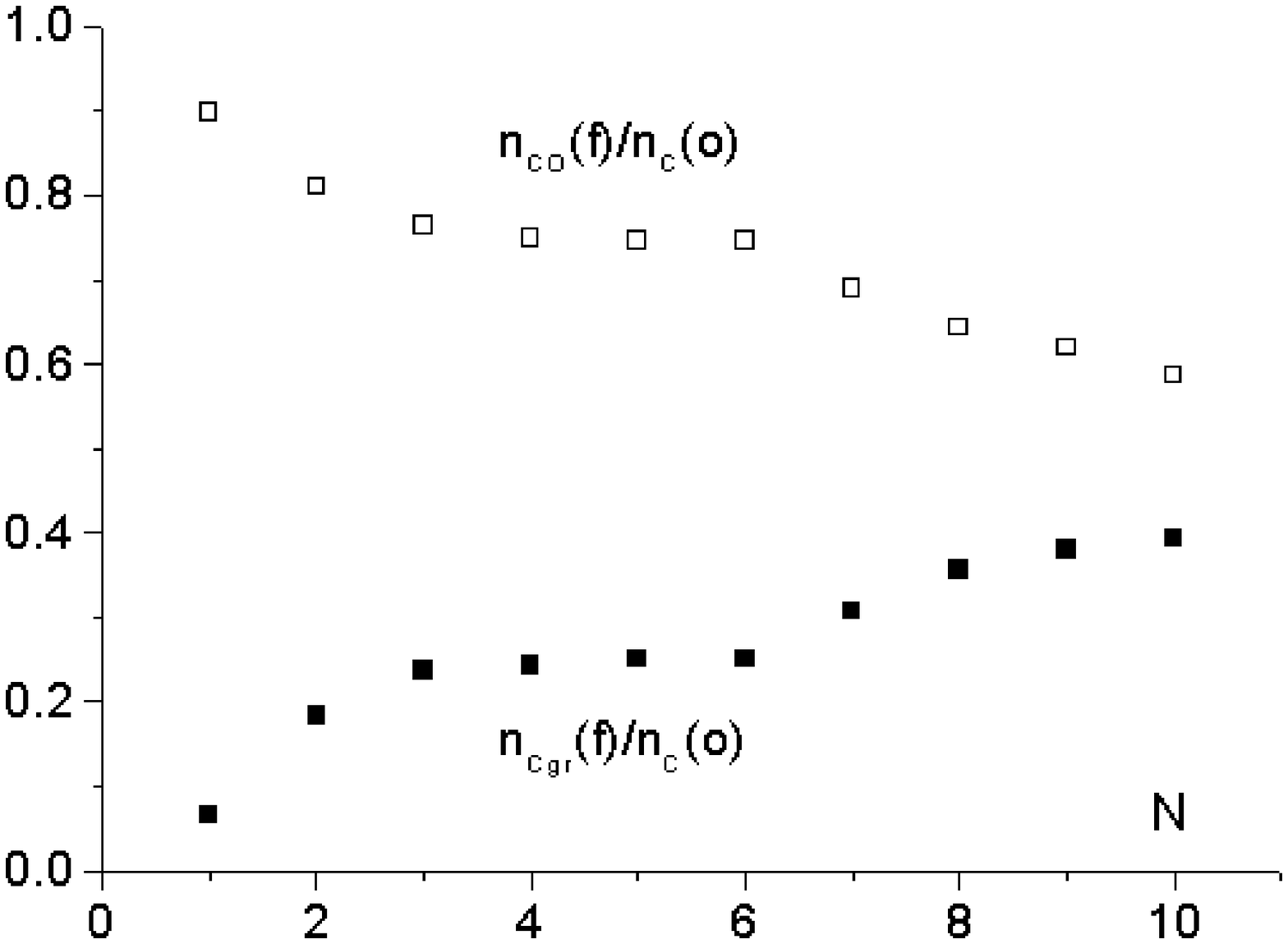}}
\caption[]{Effects of changes in the parameters, one at a time, defined by the rank, N. N=1: k1=$0.1\,\,10^{-10}$. N=2: k2=0; N=3: length of run multiplied by 2. N=4: $n_{OH}=0$ at all times. N=5: all rates as in Table 1. N=6: $n_{H2}(0)/n_{H}=0.1$. N=7: k1=$2\,\,10^{-10}$. N=8: k1=$6.6\,\,10^{-10}$.N=9: $a_{316}=0$. N=10: $a_{241}+a_{242}=0$. C/O=1 for all.}
\end{figure}

Back in Fig. 4 and 5, dotted lines have been drawn to illustrate the application of the rule of thumb recalled in the Introduction. As expected, the latter approximation is seen to agree with our more elaborate calculations only in the two opposite limits of $n_{C}(0)/n_{O}(0)$. The difference is considerable in between, which is the most often encountered range in the sky (see for instance Kwok et al. \cite{kwo}).

In a further effort to gauge the relevance of our chemical scheme, we sought to reduce the number of intermediate species to the strict minimum. Only CH was found to be indispensable for the formation of both CO and grains. The corresponding reduced scheme is
\begin{eqnarray*}
& &\dot n_{C}=-a_{63}n_{C}n_{CH}-k_{1}\,n_{C}n_{H}+a_{1}n_{H}n_{CH};\\
& &\dot n_{O}=-a_{140}n_{O}n_{CH};\\
& &\dot n_{Cgr}=2k_{2}n_{CH}n_{CH}+k_{63}n_{CH}n_{C};\\
& &\dot n_{CH}=-a_{1}n_{H}n_{CH}-a_{63}n_{C}n_{CH}-a_{140}n_{CH}n_{O}\\
& &-2k_{2}n_{CH}n_{CH}+k_{1}n_{C}n_{H};\\
& &n_{CO}=n_{O}(0)-n_{0}.\\
\end{eqnarray*}

The results are represented in Fig. 4 and 5 by open circles. These results are clearly incorrect considering their behaviour at the two ends of the X-axis. However, one notes that  $n_{CO}(f)/n_{C}(0)$ computed with this scheme is never less than one half of the value delivered by the more elaborate scheme, and the strings of squares and circles are closely similar. This is witness to the robustness of the first scheme.

This was further tested by performing other runs with the first scheme, only changing one or two parameters. Figure 7 plots $n_{CO}(f)/n_{C}(0)$ (open squares) and $n_{Cgr}(f)/n_{C}(0)$ (filled squares), for some of these cases, all with $n_{C}(0)=n_{O}(0)=3\,\,10^{6}$ cm$^{-3}$. The integer abscissa refers to the sentence in the legend which describes the corresponding changes applied to the parameters. The results are arranged so as to higlight the opposite effects of various changes in the chemical scheme, with respect to the ``standard" case, corresponding to the equations of Sec. 3.
Some elements of the scheme have no great impact on the final values of interest here; for instance, the inclusion of OH (N=3) and the initial fraction of molecular hydrogen (N=5). By contrast, the association of C and H to form CH (reaction k1) is essential for a high yield of carbon grains, as stated above and as evidenced by the steep reduction in carbon grain formation when k1 is reduced to $0.1\,\,10^{-10}$ (N=1). Other parameters have moderate impact.

Note that, in case N=1, the yield of CO has not diminished, and even increased somewhat, although CH is an important factor of CO production through reaction 140 (Table 1). This is because other quite efficient, reactions are still at play, which compensate the blocking of that route. However,  the steady state is retarded by a factor 10.

Other computations also showed that changing all initial atom densities in the same proportions by a factor 2 had only marginal effects on the final yields, but retarded or advanced the setting of steady state.

Finally, let us consider the problem of carbon grain formation. The fraction of cosmic carbon that is available for grain formation is usually equated to the difference between photospheric (cosmic) and ISM abundances of carbon, divided by the former. In our treatment, this is represented by the average of $n_{Cgr}(f)/n_{C}(0)$ over all stars. This quantity may be deduced, with the help of Fig. 5, from the measurement of the photospheric C/O ratio of a large number of stars of different types. $\emph {If the stars were evenly distributed over the range}$  0.5$<$C/O$<$2, the fraction would be $\sim$0.25. If, furthermore, the cosmic C/H ratio is assumed to be that recommended by Snow and Witt \cite{sno}, $2.25\,\,10^{-4}$, then, the available number of C atoms for carbon grains becomes 55 for $10^{6}$ H atoms.

However, this must be considered a lower limit, as it is the result of calculations with k1=$1\,\,10^{-10}$ cm$^{3}$s$^{-1}$, a reasonable but arbitrary value, adopted in Table 1, in the absence of experimental data. If, on the other hand, one takes k1=$6.6\,\,10^{-10}$, as suggested by our chemical modeling (see Appendix), and if, moreover, one adopts the Grevesse cosmic carbon abundance for the Sun \cite{gre}, i.e. 4$\,\,10^{-4}$, then  the number of available C atoms rises to 144.

  Circumstances may be envisioned where this upper limit may still be raised to some extent. For instance, as the expanding gas reaches the outskirts of the envelope, it is no longer shielded from the UV radiation of the ISM. If the latter is strong enough, CO may then be dissociated, making some more free C atoms available for grain formation, as observed by Herpin et al. \cite{her} in the case of post-AGB stars.

 Another instance is a high initial ratio of molecular to atomic hydrogen. Following a suggestion by the reviewer of this paper, our program was run (after minor changes) for the case where $n_{H}(0)=n_{H2}(0)=10^{10}$cm$^{-3}$ and $n_{C}(0)=n_{O}(0)=3\,\,10^{6}$ cm$^{-3}$. Comparing with Fig. 1 to 3, the final value of $n_{Cgr}$ was found to increase to 1.2$\,\,10^{6}$ from 0.75$\,\,10^{6}$ cm$^{-3}$, and the final value of $n_{O}$ to decrease to $1.8\,\,10^{6}$ from $2.24\,\,10^{6}$ cm$^{-3}$. This notable increase in grain abundance in the presence of a high density of H$_{2}$ is due to reactions 48 and 237 which generate CH$_{2}$ and C$_{2}$H$_{2}$, the progenitors of grains. 

\rm The range 55 to 144 for $10^{6}$ H atoms, deduced from our model, is consistent with spectroscopic observations of a couple of stars as reported by Snow and Witt \cite{sno}, in a compilation of measurements. While more astronomical observations could help, the present work should be an encouragement to further study simple chemical reactions between neutrals, especially reaction k1, so as to restrict the range of uncertainty of the model. Such reactions have become more amenable to measurement with the development of specialized apparata, e.g. the CRESU supersonic-flow machine
(see for instance Rowe \cite{row}).

\section{Appendix: Chemical modeling of binary reactions}
The rate listed in the UMIST Data Base, Rate06, for the radiative association of C and H to give CH is $10^{-17}$ cm$^{3}$s$^{-1}$. Such a low value can hardly start a chain of reactions leading ultimately to the formation of enough carbonaceous grains to fit observations. For want of other experimental data, we resorted to chemical simulation. This consists in using molecular dynamics to represent the collision of C and H atoms, for different relative velocities and impact parameters, to yield the cross-section for association as a function of relative velocity. At the end of a run, either the two atoms fly away in opposite directions, or the CH molecule is formed, with internal and external energies such that the initial energy and momentum are conserved. An order of magnitude cross-section is then deduced from these results; combined with a reference temperature (and hence a velocity of impact), this yields the rate, k.

A description of the software package used for simulation can be found in Papoular \cite{pap01}. The particular code used here is AM1 (Austin Model 1), as proposed, and optimized for carbon-rich molecules, by Dewar et al. \cite{dew85}. This semi-empirical method combines a rigorous quantum-mechanical formalism with empirical parameters obtained from comparison with experimental results. It computes approximate solutions of Schroedinger's equation for each instantaneous spatial distribution of atomic nuclei, and uses experimental data only when the Q.M. calculations are too difficult or too lengthy. This makes it more accurate than poor ab initio methods, and faster than any of the latter.

The molecular dynamics relies on the Born-Oppenheimer approximation to determine the motions of atoms under nuclear and electronic forces due to their environment. At every step, all the system parameters are memorized as snapshots so that, after completion of the run, a movie of the reactions can be viewed on the screen, and the trajectory of any atom followed all along. This makes for a better understanding of the details of  mechanisms and outcomes. In particular, it becomes possible to measure the activation energy, if any.

Note that the dynamics of bond dissociations and formation can only be simulated by using Unrestricted Hartree-Fock (UHF) wave functions in the Q.M. part of the calculation (see Szabo and Ostlund \cite{sza89}). The elementary calculation step was set at $10^{-16}$ s.

Semi-empirical simulation methods do not account for purely quantum-mechanical phenomena like tunneling or zero-point energy. In the present context of relatively high temperatures, the former is irrelevant, and the latter hardly affects the probability and energetics of the physico-chemical reactions involved. The simulation does not treat absorption or emission of light; so the contribution of line radiation to the energy and momentum budgets is not included, which is a weakness of the procedure. However, CH, for instance, has only 3 resonance lines, all of order 3 to 4 eV (Radzig and Smirnov \cite{rad}), not enough to play a significant role in the reaction budget.

The computations were performed for impact parameters p=0, 1, 2, 3, 4, 5 $\AA{\ }$ and H velocities, V, between 5 and 110 $\AA{\ }$ per picosecond ($5\,10^{4}$ and $1.1\, 10^{6}$ cm.s$^{-1}$). No association was observed for p$>$3 $\AA{\ }$, nor for V  higher than $5\, 10^{5}$ cm.s$^{-1}$ (which corresponds to a H gas temperature of $\sim$1200 K). For p=3 $\AA{\ }$, association occured for all velocities lower than $5\, 10^{5}$cm.s$^{-1}$. No activation energy was detectable.

If the cross-section radius is taken as  3 $\AA{\ }$, the area becomes $27.2\,\,10^{-16}$ cm$^{2}$. At a gas temperature of 300 K (reference temperature for rates in Rate06), the relative velocity of C and H is $2.4\,\,10^{5}$ cm.s$^{-1}$ and k1=$6.6\,\,10^{-10}$ cm$^{3}$s$^{-1}$. However, caution is in order here, because the dependence on temperature of neutral-neutral reaction rates in general is not yet fully documented experimentally (see Smith et al. \cite{smi}); \rm so k1 was set at $10^{-10}$ cm$^{3}$s$^{-1}$ in Table 1, but computations were also made for 0.1 and 6.6 $10^{-10}$ cm$^{3}$s$^{-1}$, for comparison. Similar computations were made for CH+CH$\to$C$_{2}$H$_{2}$.

\end{document}